# High-fidelity lunar topographic reconstruction across diverse terrain and illumination environments using deep learning

Hao Chen [a, *], Philipp Gläser [a], Konrad Willner [b], Jürgen Oberst [a]

[a] *Institute of Geodesy and Geoinformation Science, Technische Universität Berlin, Berlin 10553, Germany*

[b] *Institute of Planetary Research, German Aerospace Center (DLR), 12489 Berlin, Germany*

[*] Corresponding Author: hao.chen.2@campus.tu-berlin.de

**Abstract:** Topographic models are essential for characterizing planetary surfaces and for inferring underlying geological processes. Nevertheless, meter-scale topographic data remain limited, which constrains detailed planetary investigations, even for the Moon, where extensive high-resolution orbital images are available. Recent advances in deep learning (DL) exploit single-view imagery, constrained by low-resolution topography, for fast and flexible reconstruction of fine-scale topography. However, their robustness and general applicability across diverse lunar landforms and illumination conditions remain insufficiently explored. In this study, we build upon our previously proposed DL framework by incorporating a more robust scale recovery scheme and extending the model to polar regions under low solar illumination conditions. We demonstrate that, compared with single-view shape-from-shading methods, the proposed DL approach exhibits greater robustness to varying illumination and achieves more consistent and accurate topographic reconstructions. Furthermore, it reliably reconstructs topography across lunar features of diverse scales, morphologies, and geological ages. High-quality topographic models are also produced for the lunar south polar areas, including permanently shadowed regions, demonstrating the method's capability in reconstructing complex and low-illumination terrain. These findings suggest that DL-based approaches have the potential to leverage extensive lunar datasets to support advanced exploration missions and enable investigations of the Moon at unprecedented topographic resolution.

## 1. Introduction

The Moon, the most extensively studied celestial body beyond Earth, is a key reference for understanding many planetary processes (Crawford et al., 2021; Gaddis et al., 2023). Historically, our understanding of the lunar surface has been constrained by the limited resolution of available remote sensing datasets (Bickel et al., 2021; Gläser et al., 2021; Barker et al., 2023). Although high-resolution observations have been increasingly available, methodologies for fully exploiting these data have yet to mature, leaving many surface features insufficiently resolved, particularly from a three-dimensional (3D) perspective (Chen et al., 2022, 2024). For example, advances in orbital optical imaging systems enable high-resolution observations of the lunar surface at submeter to meter scales, such as those provided by the Narrow Angle Camera (NAC) onboard the Lunar Reconnaissance Orbiter (LRO) (Robinson et al., 2010). This improved spatial resolution effectively mitigates the texture homogenization of datasets coarser than 10 meters/pixel, revealing diagnostic morphological features that provide detailed insights into lunar surface processes (Henriksen et al., 2017; Barker et al., 2023). Such capabilities enable the distinction of meter-scale secondary impact clusters through ejecta pattern analysis, the delineation of volcanic vent margins from impact melt boundaries based on albedo/texture transitions, and the identification of landforms potentially indicative of subsurface volatiles, among other examples (Lawrence et al., 2013; Zhang et al., 2016; Chang et al., 2021; Neish et al., 2021; Basilevsky et al., 2025). However, although high-resolution images reveal detailed surface morphology, its inherently two-dimensional nature restricts access to quantitative topographic information, such as slope, relief, roughness, and volumetrics, required to constrain three dimensional processes governing lunar surface evolution (Oberst et al., 2014; Wu et al., 2014; Chen et al., 2025). These limitations highlight the critical role of digital terrain models (DTMs) in providing the geometric context necessary for estimating impact energies, constraining volcanic vent dynamics, analyzing tectonic structures, evaluating volatile-retention mechanisms, and interpreting a wide range of geological processes (Schleicher et al., 2019; Deutsch et al., 2020; Moon et al., 2021; Watters, 2022; Broquet & Andrews-Hanna, 2024; Zhang et al., 2024).

Traditional approaches to lunar DTM reconstruction based on image observations primarily include stereo-photogrammetry (SPG) and shape-from-shading (SFS), which have been well developed and whose performance has been extensively evaluated (Preusker et al., 2011; Scholten et al., 2012; Grumpe et al., 2014). SPG is a geometry-based technique that reconstructs terrain by directly solving imaging geometry from stereo parallax measurements, enabling accurate elevation reconstruction when the stereo configuration is well constrained (Oberst et al., 2014; Henriksen et al., 2017). In lunar applications, its performance can be constrained by limited stereo overlap, suboptimal convergence angles, and areas of insufficient image texture, potentially resulting in gaps or reduced accuracy in the reconstructed DTMs (Liu & Wu, 2020; Chen et al., 2022). SFS is a photometry-based technique that reconstructs topography by inferring surface elevations from image shading under known illumination conditions, enabling high-fidelity DTM reconstruction from a single image (Wu et al., 2018). Its performance, however, is sensitive to variations in illumination and surface reflectance, which can result in local inconsistencies in the reconstructed DTMs (Alexandrov & Beyer, 2018). SFS can also be applied using multiple images under different illumination conditions to improve reconstruction, but this strategy increases the complexity of image selection and processing, and is typically applied to local areas, such as landing sites (Boatwright & Head, 2024). In recent years, deep learning (DL)-based methods have rapidly advanced, enabling high-resolution topographic modeling of planetary surfaces from single-view images (La Grassa et al., 2022; Liu et al., 2022; Tao et al., 2023). However, DL-based methods often produce discrepancies and artifacts when compared with reference models (e.g., from SPG) and generally achieve lower effective resolution than SFS methods, resulting in compromised reconstruction quality. To address these limitations, our team developed a novel DL-based single-view method, termed ELunarDTMNet, which incorporates coarse-resolution topographic constraints to improve lunar DTM reconstruction (Chen et al., 2024). The method resolves small-scale topographic details comparable to SFS approaches, while reproducing overall terrain morphology consistent with SPG-derived DTMs, and maintains substantially faster processing speed, demonstrating the potential for efficient, high-quality topographic mapping.

The lunar surface exhibits a diverse range of terrain features, from large-scale structures such as mountains and impact basins to subtle features such as ejecta rays and boulders. This diversity encompasses both regular structures, such as simple craters, and irregular forms, including lava flows and wrinkle ridges, reflecting geological processes spanning from ancient to more recent periods. An efficient topographic modeling method is therefore essential to reconstruct the Moon's surface with sufficient fidelity to support reliable morphological analysis across diverse landforms. Moreover, variations in solar illumination introduce variable shading and reflectance effects, which challenge the consistency and geometric reliability of terrain information extracted from optical imagery for topographic mapping (Preusker et al., 2015; Liu & Wu, 2021). Robust performance across diverse terrain types and varying illumination conditions is a prerequisite for high-quality, large-scale topographic mapping; however, questions regarding the robustness and consistency of DL-based approaches under such conditions remain open in planetary applications. In this study, we use improved ELunarDTMNet as a representative DL framework to systematically assess the reliability capability of DL-based topographic reconstruction, with particular emphasis on performance under varying solar incidence angles and across terrain features of differing morphology and geological age. In addition, we extend our assessment to the lunar south pole with permanently shadowed regions (PSRs), which represent challenging yet scientifically important targets for future lunar exploration. As these regions lie outside the mid- to low-latitude domains for which ELunarDTMNet was originally developed, we adapt the network through fine-tuning for low-illumination environments. The goal of this study is to evaluate whether DL-based methods can serve as a reliable and complementary approach for high-fidelity lunar topographic reconstruction, augmenting established techniques such as laser altimetry, SPG, and SFS, particularly in regions where constraints in data coverage, spatial resolution, and imaging conditions limit the applicability of traditional methods.

## 2. Data and Methodology

### 2.1. Input data: high-resolution images and coarse-resolution DTMs

ELunarDTMNet comprises two complementary sub-networks: a high-resolution image branch that encodes high-frequency terrain features, and a coarse-resolution DTM branch that imposes low-frequency elevation constraints (Chen et al., 2024). Accordingly, the network requires a high-resolution image together with a coarse-resolution DTM as inputs for topographic modeling.

#### 2.1.1. LRO NAC and ShadowCam Observations

The LRO mission hosts two NAC subsystems, NAC-Left (NAC-L) and NAC-Right (NAC-R), which capture high-resolution panchromatic images of the lunar surface (Robinson et al., 2010). The NAC subsystems are mounted to image adjacent swaths simultaneously, achieving a broad coverage while maintaining a pixel scale of 0.5-2 m, providing fine-scale terrain details essential for high-resolution DTM reconstruction (Haase et al., 2012; Henriksen et al., 2017). To obtain stereo observations, the spacecraft will need to be especially pointed off-nadir within the subsequent orbit. ELunarDTMNet was originally trained on NAC-derived datasets to model the relationship between imagery and underlying lunar terrain morphology (Chen et al., 2024). Although NAC stereo coverage is spatially sparse, high-resolution single-view NAC images cover nearly the entire Moon, providing a basis for assessing ELunarDTMNet performance across varied landscapes.

In the polar regions, low solar elevation angles (high incidence angles) result in PSRs within craters and depressions, where the retrieval of surface morphological details from conventional NAC optical imagery is ineffective (Mazarico et al., 2011; Robinson et al., 2023). ShadowCam, a NASA-funded instrument aboard the Korean Pathfinder Lunar Orbiter (KPLO), overcomes this limitation with a sensitivity over 200 times greater than that of the NACs (Robinson et al., 2023). Operating at a pixel scale of ~1.7 m from a nominal 100 km orbit, it resolves fine-scale features within PSRs (Speyerer et al., 2024). ShadowCam imagery thus provides an additional testbed for evaluating ELunarDTMNet's generalization across diverse imaging conditions, particularly in PSRs not well covered by NAC observations.

#### 2.1.2. SLDEM and LOLA DTM

Currently available low-resolution DTMs of the Moon, provide near global coverage and are primarily derived from laser altimetry measurements and stereo-photogrammetric reconstructions (Scholten et al., 2012; Smith et al., 2017). The mid- to low-latitude within ±60° are covered by photogrammetric DTMs produced from the Selenological and Engineering Explorer (SELENE/Kaguya) mission at a resolution of 512 pixels per degree (~60 m), co-registered with Lunar Orbiter Laser Altimeter (LOLA) tracks, constituting the widely used SELENE Lunar Digital Elevation Model (SLDEM) (Barker et al., 2016). These DTMs exhibit high quality for general terrain morphology and are adopted in this study as the coarse-resolution input for mid- to low-latitude regions. In the lunar polar regions, the polar orbit of LRO results in a higher density of LOLA tracks compared with low-latitude regions, providing denser elevation sampling for gridded DTM generation (Gläser et al., 2014). Barker et al. (2021, 2023) further refined these LOLA-derived DTMs by mitigating artifacts caused by track geolocation errors, producing high-quality gridded models, for example, with 20 m resolution poleward of 80°S. These improved polar DTMs are used as coarse-resolution inputs, complementing high-resolution NAC or ShadowCam imagery, for producing high-resolution DTMs in the south polar region.

**2.2. ELunarDTMNet-based high-resolution topographic modeling**

In this study, we use ELunarDTMNet for lunar DTM reconstruction (Chen et al., 2024), incorporating an improved scale-recovery strategy and a fine-tuned model for low solar elevation environments, building on our developed large-area topographic mapping framework (Chen et al., 2022). We first evaluate the effect of varying solar incidence angles on reconstruction performance (Sect. 3.1), and then employ the framework to reconstruct topography across diverse lunar terrain types to assess robustness and generalization (Sect. 3.2), including the south polar region (Sect. 3.3). The image datasets used are listed in Supplementary Tables S1 and S2.

DL networks typically process DTMs scaled to a normalized range; therefore, the predicted outputs need to be rescaled to absolute elevations, usually with reference to a coarse-resolution model such as LOLA-derived topographic models (Gläser et al., 2014; Chen et al., 2022; Tao et al., 2023). Previously, ELunarDTMNet rescaled elevations via a global max-min mapping, which, while simple to process, is sensitive to extreme values, making it vulnerable to noise and outliers and susceptible to bias from pronounced local terrain variations. In this work, we replace this strategy with a robust RANSAC-based scale recovery. Specifically, absolute elevations are obtained using a linear model

$$DTM_{abs} = aDTM_{pred} + b, \tag{1}$$

where $DTM_{pred}$ and $DTM_{asb}$ denote the predicted normalized DTM and the DTM with recovered absolute scale, respectively. The scale *a* and offset *b* are estimated by RANSAC from elevation correspondences between the predicted DTM and a coarse-resolution reference DTM. By fitting the model using the dominant set of consistent samples, RANSAC effectively avoids the limitations associated with the max-min mapping scheme and yields a more robust scale estimation. The performance comparison is provided in Table 1, where root mean square errors (RMSEs) of absolute elevation differences are reported for ELunarDTMNet under the max-min and RANSAC-based scale recovery strategies, using SPG-derived DTMs as the reference (Henriksen et al., 2017).

Table 1. RMSE of absolute elevation differences for ELunarDTMNet under different scale recovery strategies, with SPG-derived DTMs as reference.

| Test area | Max-min | RANSAC |
| --- | --- | --- |
| Hponds | 2.67 m | **2.46 m** |
| Apollo11 (M1149338748RE) | 2.80 m | **2.55 m** |
| Apollo11 (M1282310415LE) | 2.69 m | **2.20 m** |
| Apollo11 (M1114014396RE) | 2.61 m | **2.19 m** |
| Apollo11 (M104362199RE) | 2.53 m | **2.17 m** |
| Apollo11 (M150361817RE) | 2.58 m | **2.13 m** |
| Apollo11 (M1157600009RE) | 2.61 m | **2.09 m** |
| Lunar south polar region (poleward of -88.5°) | 0.92 m | **0.65 m** |

Note: Values with higher accuracy are highlighted in **bold**. The Hponds test area is shown in Chen et al. (2024). The Apollo 11 test area is shown on Fig. 1. The lunar south polar test area is shown on Fig. 7.

ELunarDTMNet is trained using high-quality SPG-derived NAC DTMs as ground truth. The original training dataset is restricted to latitudes within ±60°, using NAC images together with SLDEM as network inputs. As a result, the original model is primarily adapted to mid- to low-latitude illumination and topographic conditions. In contrast, the polar regions experience persistently low solar elevation angles, whereas low-latitude regions are exposed to higher sun angles, resulting in markedly different illumination conditions. In addition, polar DTMs are typically generated from denser LOLA track coverage, producing gridded models with distinct spatial characteristics compared with SLDEM. These differences introduce a substantial domain shift when applying the original model to polar regions. Here, we fine-tune ELunarDTMNet using SPG-derived polar regional DTMs released through the LROC Planetary Data System (PDS) (Henriksen et al., 2017), with NAC images and 20 m resolution polar LOLA DTMs as inputs, in order to adapt the network to the unique terrain and illumination characteristics of the polar regions. Specifically, following Chen et al. (2022, 2024), we construct the training and validation datasets by splitting the original images and DTMs into overlapping patches, with an overlap of 100 pixels between adjacent patches. Data augmentation is performed by applying random horizontal and vertical flips. The model trained on mid- to low-latitude data serves as a pre-trained initialization and is subsequently fine-tuned using polar region datasets. Fine-tuning is conducted using a small initial learning rate of 0.00001 over 20,000 training iterations. Model validation is performed every 5,000 iterations, and the network checkpoint yielding the best validation accuracy is selected as the well-trained model for polar topography inference.

# 3. Results

## 3.1. Reconstruction performance under varying solar illuminations

Solar azimuth and incidence angles critically affect the quality and consistency of image-based topographic reconstruction, especially for single-view approaches. The high-resolution LRO NAC images are dominated by solar illumination along a principal direction, so the brightness gradients provide strong normal constraints along the illumination direction, while offering only limited information in the orthogonal direction. Because single-view SFS infers surface slopes from brightness variations, strong gradients along the illumination direction provide reliable constraints, whereas weak gradients perpendicular to the sunlight render height estimates under-constrained, reducing reconstruction accuracy (Alexandrov & Beyer, 2018; Chen et al., 2022). Previous studies have shown that DL-based methods can produce stable DTMs from LRO NAC images acquired under varying solar azimuths, yielding consistent reconstructions both along and across the illumination direction (Chen et al., 2022). On the other hand, the solar incidence angle has a direct impact on image intensity via the interaction between incoming light and surface reflectance. Robust interpretation of these incidence-angle-driven reflectance variations is essential for ensuring model accuracy. The performance of SFS can be affected by changes in solar incidence angle, as such variations alter the observed brightness and reduce the reliability of topography cues (Liu & Wu, 2021). When the illumination geometry deviates from the optimal range, these cues become less informative for shape recovery. In this section, we investigate the influence of solar incidence angle on ELunarDTMNet and assess its performance in comparison with the SFS method (Alexandrov & Beyer, 2018). Six LRO NAC images at 1.5 m spatial resolution, with solar incidence angles ranging from approximately 10° to 70° and all covering the same Apollo11 landing area, are used for testing.

Figure 1a shows that increasing solar incidence angles are associated with reduced overall brightness and enhanced shadow contrast in the NAC images, which improves the discernibility of surface morphology. The ELunarDTMNet-derived 1.5 m resolution DTMs in Fig. 1b exhibit overall consistency, particularly for incidence angles larger than 30°. By comparison, the DTM generated from the low incidence angle of 13.1° shows differences relative to those derived from higher incidence angles. For craters with diameters on the order of hundreds of meters, where brightness does not fully obscure topography, the elevation profiles (profiles 1 and 2 in Fig. 1c) indicate that the overall crater morphology is preserved across different incidence angles. In particular, the bowl-shaped geometry is evident even for the incidence angle of 13.1°, similar to that observed at higher incidence angles. For craters with diameters on the order of tens of meters, which are sometimes more strongly affected by brightness effects, the elevation profiles (profiles 3 and 4 in Fig. 1c) show larger discrepancies at the low incidence angle of 13.1°. In these cases, high surface brightness obscures small-scale topographic features in the NAC images, making the craters less distinguishable compared to images acquired at higher incidence angles. This behavior is analogous to known limitations in single-view SFS methods: under low solar incidence angles, the images appear relatively bright and albedo-dominated, which hinders the accurate reconstruction of surface topography (Liu & Wu, 2021).

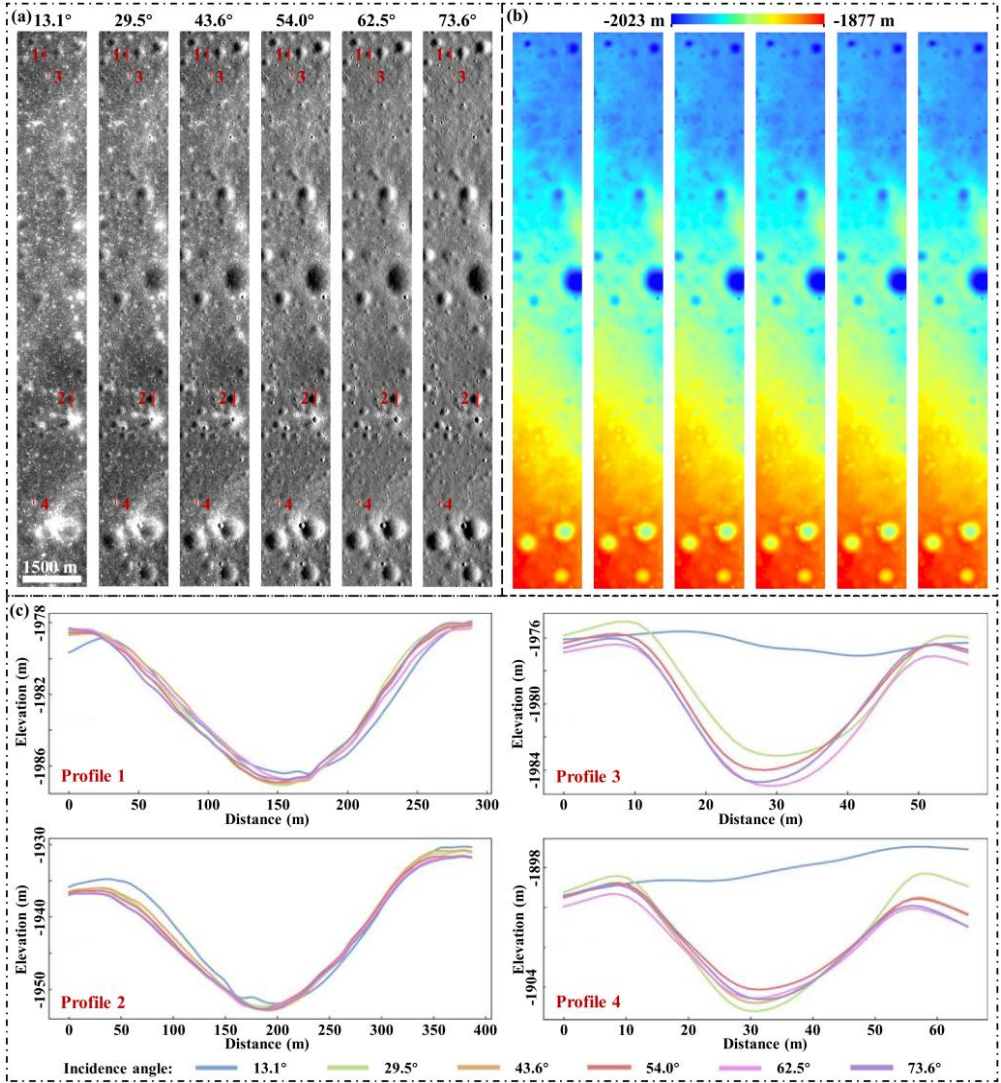

Fig. 1. Topographic reconstruction results under varying solar incidence angles. (a) LRO NAC images at 1.5 m resolution of the same Apollo11 landing area, acquired at solar incidence angles of 13.1° (M1149338748RE), 29.5° (M1282310415LE), 43.6° (M1114014396RE), 54.0° (M104362199RE), 62.5° (M150361817RE), and 73.6° (M1157600009RE). (b) Corresponding 1.5 m resolution DTMs generated using ELunarDTMNet. (c) Elevation profiles of four craters derived from the DTMs: two with diameters on the order of hundreds of meters and two on the order of tens of meters. The red lines in (a) indicate the profile locations.

We further assess the performance of ELunarDTMNet and SFS by applying both methods to the same input. For ELunarDTMNet-derived DTMs, large errors generally occur in areas with significant topographic variations. In contrast, SFS-derived DTMs not only exhibit large errors in regions with strong topographic variations, but also show larger errors than ELunarDTMNet in gentle terrain, as shown in Supplementary Fig. 1. Figure 2 shows RMSE values derived from the absolute elevation errors of DTMs, referenced to the SPG model. Across all solar incidence angles, ELunarDTMNet consistently achieves lower RMSE values than SFS, indicating more accurate reconstruction. As the incidence angle increases, RMSE decreases for both methods; however, ELunarDTMNet not only maintains lower errors but also exhibits a more stable trend, suggesting reduced sensitivity to variations in illumination. This consistent accuracy allows for more flexible image selection, making ELunarDTMNet better suited for large-scale terrain reconstruction while minimizing uncertainties associated with reconstruction errors.

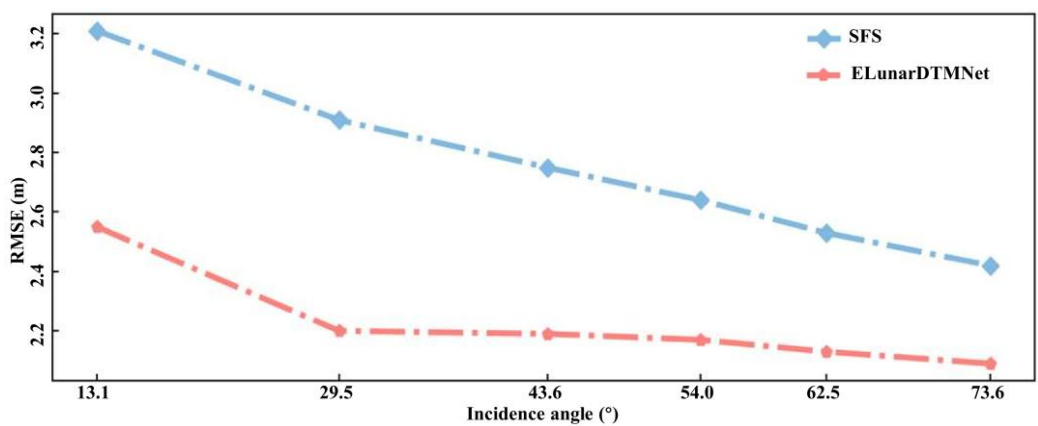

Fig. 2. Reconstruction performance comparison between ELunarDTMNet and SFS (Alexandrov & Beyer, 2018) under different solar incidence angles, quantified using RMSE derived from absolute elevation error maps and referenced to the SPG model (Henriksen et al., 2017).

### 3.2. Reconstruction results with diverse terrain features

This section examines the results of ELunarDTMNet over multiple characteristic lunar terrain types, focusing on its behavior under varying geomorphological settings. For large-scale areas of the lunar surface, topographic modeling typically requires multiple LRO NAC images to provide sufficient spatial coverage. Figure 3 shows two validation regions reconstructed using ELunarDTMNet: one with 32 NAC images (Fig. 3a), covering highlands and impact melt areas, and another with 16 NAC images (Fig. 3b), covering a region centered on a central peak. The ELunarDTMNet-derived DTMs show elevation patterns similar to the reference SPG DTM (top panel of Figs. 3a and 3b). In the local shaded models (bottom panel of Fig. 3), the higher-resolution ELunarDTMNet DTMs capture more comprehensive terrain details compared with the SPG DTM. In small, shaded regions where the SPG DTM contains gaps, ELunarDTMNet provides a complete terrain representation, as indicated by the yellow arrows in Fig. 3b. More large-scale topographic reconstructions are provided in Supplementary Figs. S2, S3, and S4, illustrating key terrain features, including mare regions, volcanic structures, etc. Supplementary Figs. S2 and S3 show that the SPG

DTM cannot reconstruct certain regions due to the lack of stereo pairs; in contrast, ELunarDTMNet provides complete terrain representations in these areas using monocular reconstruction. In addition, reconstruction artifacts are present in the SPG DTMs, such as box-shaped features and seamlines (Supplementary Figs. S3-S5). These artifacts arise from the limitations of stereo matching in regions with low texture or repetitive patterns, as well as minor geometric misalignments between overlapping images, which can produce discontinuities at tile boundaries. In contrast, the ELunarDTMNet-derived DTMs are free of such artifacts and remain consistent with the surface morphology observed in the original NAC images, with high fidelity and seamless coverage over large lunar regions.

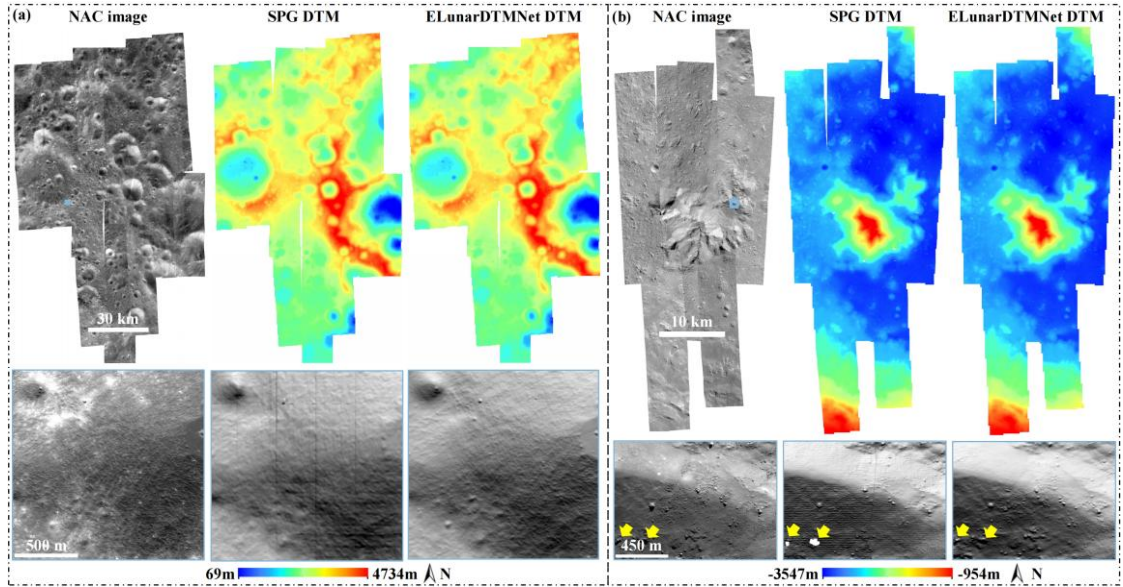

Fig. 3. Reconstruction results over two large lunar regions using multiple overlapping LRO NAC images. Top panel in (a) shows 32 NAC images at 1.7 m resolution, covering 165.2°-169.4° longitude and 39.1°-44.5° latitude, 5 m resolution SPG DTM (Henriksen et al., 2017), and 1.7 m resolution ELunarDTMNet-derived DTM. Top panel in (b) shows 16 NAC images at 1.5 m resolution, covering 348.1°-349.2° longitude and -44.4° to -42.3° latitude, 2 m resolution SPG DTM (Henriksen et al., 2017), and 1.5 m resolution ELunarDTMNet-derived DTM. The bottom panels in (a) and (b) show the corresponding local NAC images and shaded models derived from the SPG and ELunarDTMNet DTMs. These areas are indicated by blue boxes in the top panel.

Lunar craters dominate the lunar surface and serve as natural laboratories for investigating impact processes, crustal deformation, and morphological evolution, spanning a wide spectrum from simple bowl-shaped pits to complex, multi-ring structures (Osinski et al., 2023). Figure 4a presents a high-resolution topographic model of a concentric, donut-shaped crater, which may have been formed by double impacts, volcanic processes, and compositional variations. Figure 4b shows a simple, bowl-shaped crater that is among the deepest on the Moon, providing exposures of subsurface layers and making it an ideal target for investigating internal structures. As no publicly available NAC-derived SPG DTMs exist for these two regions, comparisons are instead made between the 60 m resolution SLDEM and the 1.5 m resolution ELunarDTMNet-derived DTMs. Figure 4a demonstrates that, although the SLDEM is able to resolve craters with diameters of several hundred meters, it only provides a coarse representation of their overall morphology, as illustrated by the craters outlined by the yellow circles. In contrast, the ELunarDTMNet-derived DTM reconstructs these craters with substantially greater morphological detail, faithfully resolving their fine-scale topographic features. The bottom panel of Fig. 4b highlights a localized area on the crater floor characterized by subdued topographic relief. In the SLDEM shaded model, the subtle morphology of this low-relief surface cannot be reliably interpreted. By comparison, the ELunarDTMNet-derived shaded model reveals weak but coherent topographic expressions, including low-amplitude, spatially coherent micro-topographic textures (yellow ellipses), likely associated with mass-wasting processes on the crater floor. In addition, tectonic features represent a diverse and scientifically significant class of lunar landforms, arising from compressive and extensional stresses in the crust, and offering valuable information on the Moon's structural evolution and tectonic history (Watters, 2022). Figure 4c presents the topographic reconstruction results of a lobate scarp, a compressional tectonic landform formed in response to crustal shortening, and compares the 5 m resolution SPG-derived DTM with the 1.5 m resolution ELunarDTMNet-derived DTM. The local shaded model from SPG DTM is able to reveal the general expression of the lobate scarp; however, its morphological continuity and fine-scale structure are partially obscured by noise, limiting detailed interpretation. In contrast, the ELunarDTMNet-derived shaded model clearly delineates the scarp geometry, preserving its sharpness and continuity with substantially reduced noise, and thereby enables a more reliable characterization of the tectonic landform. Furthermore, lunar domes are mound-like volcanic landforms that record the Moon's volcanic and magmatic processes, with variations in their morphology providing important constraints on eruption conditions and subsurface structure (Smith, 1973). Figure 4d presents a high-resolution DTM of a dome, revealing its topographic relief. Compared with the 3 m resolution SPG-derived DTM, the 1.5 m resolution ELunarDTMNet-derived DTM resolves finer-scale morphological variations, enabling more precise characterization of the dome's geometry and slope distribution, and facilitating quantitative analyses of lunar volcanic processes. More topographic reconstructions across a wide range of representative lunar landforms, including crater clusters, pit chains, vents, wrinkle ridges and crosscutting structures, rilles, nested rilles, and domes with summit depressions, are presented in Supplementary Figs. S6-S12. These landforms are central to investigations of impact chronology, volcanic activity, tectonic deformation, and surface-subsurface interactions on the Moon. Collectively, these results demonstrate the robust generalization capability of ELunarDTMNet across diverse geomorphological settings and highlight its role as a complementary approach to existing SPG DTMs and SLDEM, enabling more detailed and systematic investigations of lunar surface morphology at finer spatial scales.

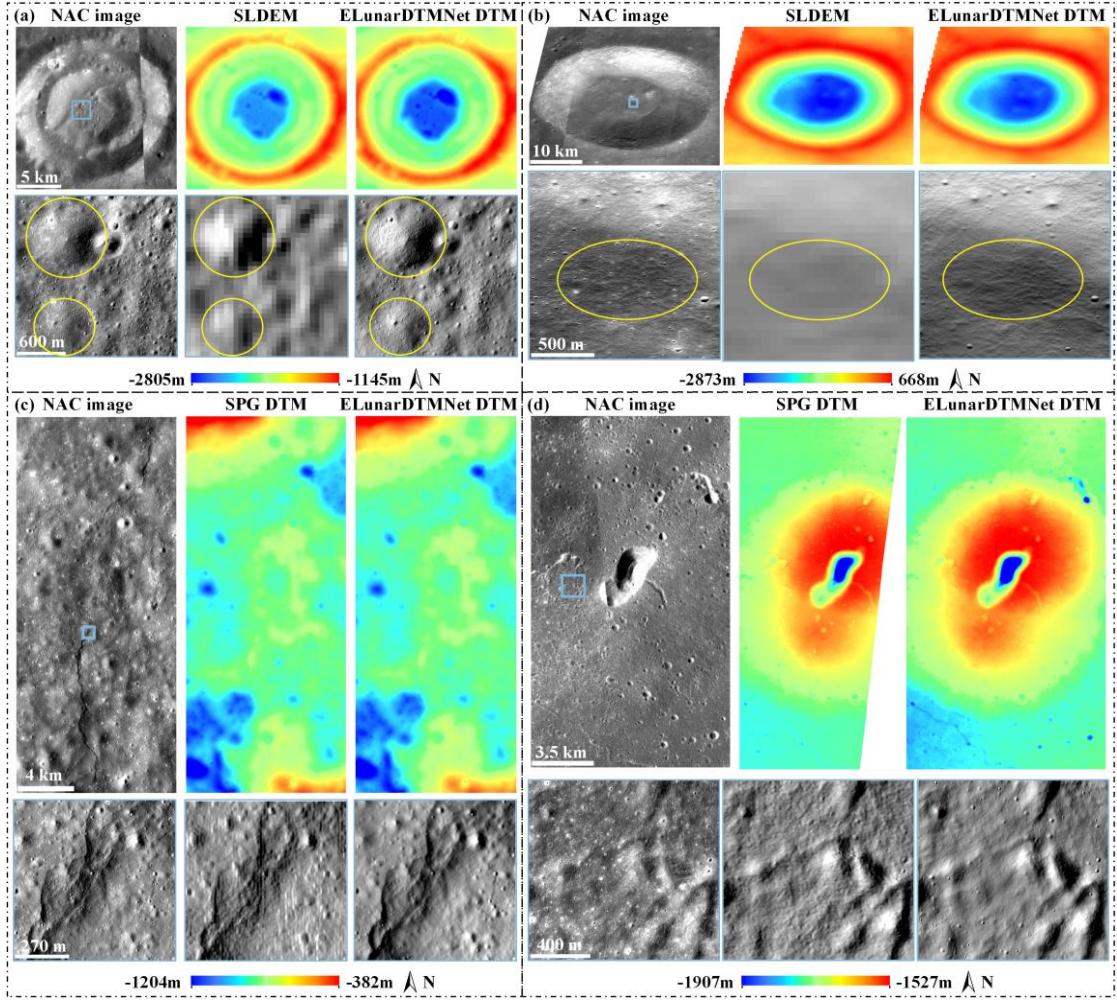

Fig. 4. Reconstruction results for four representative lunar terrain features. Subfigures show (a) the donut-shaped Bell E Crater (centered at 264°, 22°), (b) one of the Moon's deepest simple craters (centered at 267°, 52°), (c) a lobate scarp (centered at 161°, 7°), and (d) a dome (centered at 310°, 20°). In each subfigure, the top panel presents the 1.5 m resolution LRO NAC image, the reference SPG DTM (Henriksen et al., 2017) (5 m resolution for (c) and 3 m resolution for (d)), supplemented by 60 m resolution SLDEM (Barker et al., 2016) where SPG data are unavailable, together with the 1.5 m resolution ELunarDTMNet-derived DTM. The bottom panels show the corresponding local NAC images and shaded models derived from the SPG and ELunarDTMNet DTMs for regions outlined by the blue box in the top panel.

In addition to the diverse geomorphological features discussed above, geologically young landforms provide critical insights into recent volcanic and impact processes on the Moon and demand fine-scale topographic modeling to resolve their subtle morphological details. Figure 5a illustrates a geologically young, potentially very recent volcanic feature, while Fig. 5b shows a fresh impact crater with associated subtle ejecta rays. Due to the relatively coarse resolution of SPG-derived DTMs, the irregular mare patch (IMP), a manifestation of volcanism, in Fig. 5a appears blurred, and certain morphological features, as indicated by the yellow circles, are lost in the zoomed-in shaded model. In contrast, the 1.5 m resolution ELunarDTMNet-derived DTMs sharply delineate the shape of the IMP, closely matching the original NAC image. Boulders and small-scale craters near the IMP, marked by yellow arrows, are also resolved, whereas they are absent in the SPG-derived DTM. Similarly, Fig. 5b shows that, although the SPG shaded model resolves small craters and short structural features, it fails to reveal the faint ejecta rays. By comparison, the ELunarDTMNet-derived DTM not only renders these features with higher fidelity, but also reconstructs the ejecta rays, as highlighted by the yellow arrows, enabling more precise characterization of surface morphology and providing a stronger foundation for investigating recent volcanic and impact processes. Additional examples of subtle young terrain features, including boulder tracks (Supplementary Fig. S7), boulder fields (Supplementary Fig. S10), lava flows (Supplementary Fig. S13), and ejecta rays (Supplementary Fig. S14), further demonstrate ELunarDTMNet's ability to accurately capture fine-scale morphology, providing additional validation of its performance for geologically young lunar surfaces.

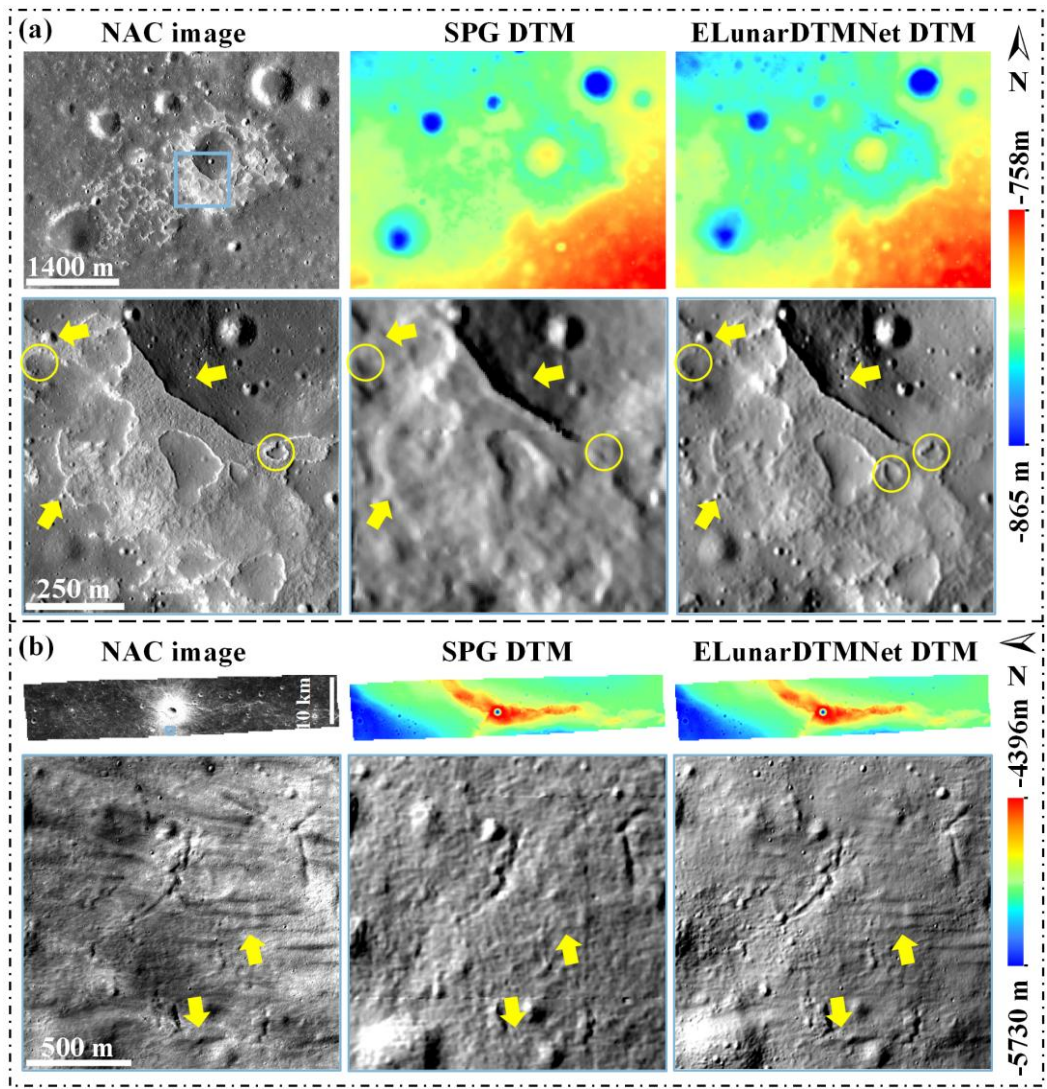

Fig. 5. Reconstruction results for two geologically young terrain features: (a) a young volcanic feature centered at (34°, 4°) and (b) a fresh impact crater centered at (25°, 30°). For each subfigure, the top panel shows the 1.5 m resolution LRO NAC image, the reference SPG DTM (Henriksen et al., 2017) (5 m resolution in (a) and 4 m resolution in (b)), and the 1.5 m resolution ELunarDTMNet-derived DTM. The bottom panel shows the corresponding local NAC image and shaded models for the area indicated by the blue box in the top panel.

### 3.3. Reconstruction results in the south polar region

In this section, the ELunarDTMNet model fine-tuned for polar conditions is applied to reconstruct topography in the south polar region. Figure 6 shows reconstruction results for a local area of Malapert Massif. The fine-tuned ELunarDTMNet-derived shaded model preserves surface details comparable to those recovered by the SFS method, which is renowned for resolving fine-scale topography, while also maintaining consistency with the original NAC image and the SPG-derived topography in overall terrain morphology. As quantified in Fig. 6e, fine-tuning ELunarDTMNet with the polar training set substantially reduces its RMSE with respect to the SPG model, from 1.37 m to 0.50 m. This value is slightly lower than that of the SFS method (0.63 m), demonstrating ELunarDTMNet's ability to produce high-quality topographic models in polar regions. Furthermore, we use the 1 m resolution LRO NAC mosaic and the 20 m resolution LOLA DTM to generate a 1 m resolution DTM for regions poleward of 88.5°S, as shown in Fig. 7a. Using the 5 m resolution LOLA DTM as a reference, fine-tuning reduces the RMSE from 1.05 m to 0.65 m. For the fine-tuned model, 84.6% of the absolute elevation errors are below 1 m and 97.6% are below 2 m, indicating good agreement with the LOLA DTM. The zoomed-in shaded models in Figs. 7b and 7c indicate that the ELunarDTMNet-derived DTM captures finer surface details than the LOLA-derived model, resolving small-scale craters and subtle surface textures that may be associated with mass-wasting processes. Although the 5 m resolution LOLA DTM is generated through interpolation to a finer grid, it shows increased track-related artifacts and reduced surface smoothness compared to the 20 m resolution LOLA DTM, thereby highlighting the complementary role of ELunarDTMNet in recovering local topographic details beyond the effective resolution of LOLA measurements. The elevation profiles in Figs. 7d and 7e show that the 1 m resolution ELunarDTMNet-derived DTM reproduces elevation trends consistent with those of the LOLA DTMs. Terrain features captured by both the ELunarDTMNet-derived and LOLA DTMs show similar elevation undulations. Differences are mainly observed in local features, which are resolved only by the ELunarDTMNet-derived DTM.

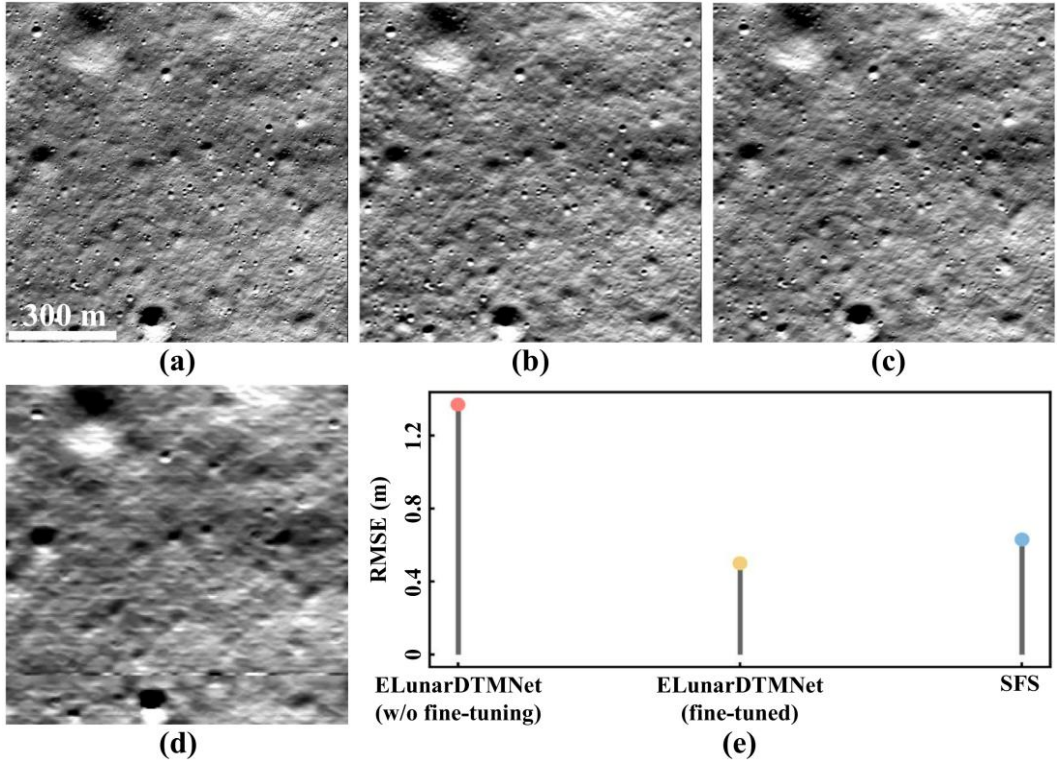

Fig. 6. Reconstruction performance in a local region of the Malapert Massif centered at (354.2°E, 85.1°S). (a) 1 m resolution LRO NAC image. (b) 1 m resolution shaded model derived from fine-tuned ELunarDTMNet DTM. (c) 1 m resolution shaded model derived from SFS DTM (Alexandrov & Beyer, 2018). (d) 4 m resolution model derived from SPG DTM (Henriksen et al., 2017). (e) RMSE quantified from absolute elevation error maps of ELunarDTMNet (without and with fine-tuning) and SFS DTMs, using the SPG DTM as reference.

Nevertheless, LRO NAC images are insensitive in PSRs due to the lack of illumination, which hinders the retrieval of detailed terrain information, as shown in Supplementary Fig. S15. To complement the NAC-based reconstruction, 2 m resolution ShadowCam images and the 20 m resolution LOLA DTM are used to generate 2 m resolution DTMs for PSRs based on the ELunarDTMNet method. These ShadowCam-derived DTMs are then merged with NAC-derived DTMs in illuminated areas to produce continuous 2 m resolution coverage across both illuminated and shadowed regions. Figure 8 presents the reconstructions of a large PSR (Shackleton crater) and a small PSR near Shackleton crater, with the NAC and ShadowCam images shown in Fig. 8a. Figure 8b shows the ELunarDTMNet-derived shaded models of the target PSRs and elevation profiles across the overlapping area between the illuminated regions imaged by NAC and the shadowed regions imaged by ShadowCam. The seamless transitions observed in the shaded models, together with the continuous elevation profiles, demonstrate the effectiveness of the DTM fusion and the consistency of terrain reconstruction across both illuminated and shadowed areas. Figure 8c presents a performance comparison between the ELunarDTMNet-derived DTM and the LOLA DTMs, focusing on a local region within the large Shackleton PSR. LOLA, as an active laser altimeter, provides reliable topographic measurements within PSRs independent of solar illumination; however, its limited footprint spacing and sampling density restrict the resolution of small-scale local terrain features. As a result, small-scale craters and boulders are resolved in the ELunarDTMNet results derived from ShadowCam images but are absent in the LOLA-derived shaded models. Elevation profiles further indicate that the overall trends of the ELunarDTMNet-derived DTM are consistent with those of the LOLA DTMs, with differences primarily limited to small-scale local terrain features, such as boulders in Fig. 8c, which are uniquely resolved by ELunarDTMNet. In this context, LOLA DTM serves as a reliable elevation constraint and absolute scale reference for ELunarDTMNet, forming a complementary framework in which ELunarDTMNet enriches topographic representation by resolving fine-scale terrain details beyond the intrinsic resolution of laser altimetry.

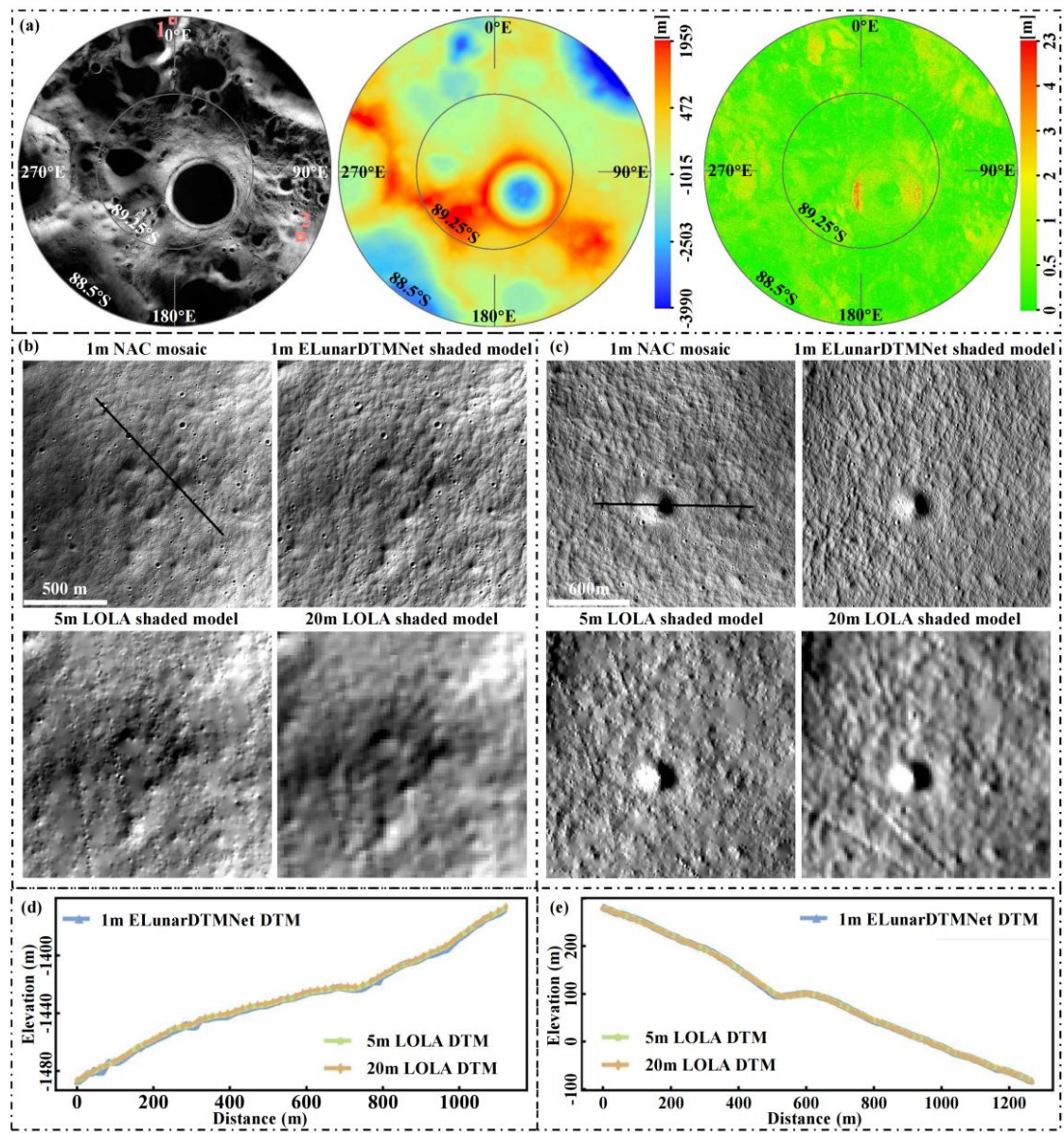

Fig. 7. Reconstruction performance for regions poleward of 88.5°S. (a) The 1 m LRO NAC mosaic (Archinal et al., 2023), the 1 m ELunarDTMNet-derived DTM, and the absolute elevation error map

of the ELunarDTMNet-derived DTM with respect to the 5 m LOLA DTM (Barker et al., 2023). (b, c) Local Area 1 and Local Area 2 comparisons, showing the 1 m NAC mosaic, the 1 m ELunarDTMNet-derived shaded model, and the 5 m and 20 m LOLA-derived shaded models; the locations of the two areas are indicated by the pink boxes in (a). (d) and (e) Elevation profile comparisons among the 1 m ELunarDTMNet-derived DTM and the 5 m and 20 m LOLA DTMs along the black profiles marked in (b) and (c), respectively.

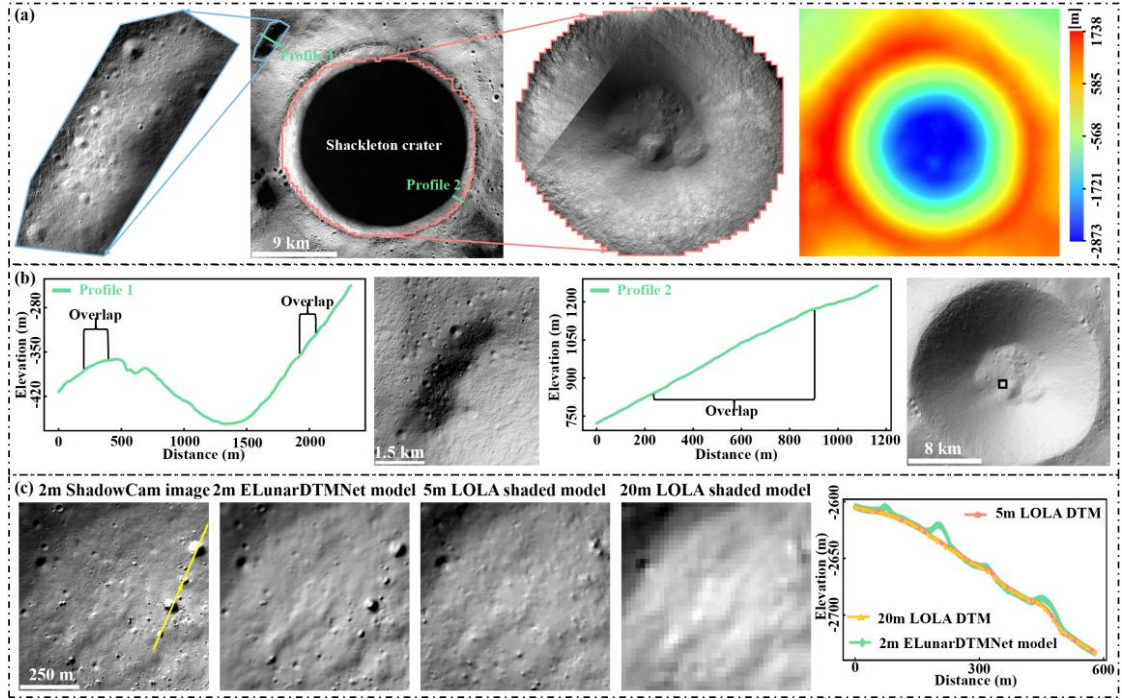

Fig. 8. Reconstruction performance for a small PSR and a large PSR (Shackleton crater). (a) LRO NAC mosaic image (Archinal et al., 2023) showing the locations of the PSRs, overlaid with 2 m resolution ShadowCam images, and the derived merged 2 m resolution DTM. (b) Shaded models of the small and large PSRs, together with elevation profiles across the overlap region between NAC and ShadowCam images, showing the seamless and continuous transition across the overlap region. The profile locations are indicated by green lines in panel (a). (c) The 2 m resolution ShadowCam image and the corresponding 2 m ELunarDTMNet shaded model, together with the 5 m and 20 m LOLA shaded models (Barker et al., 2023), for a local area within the large PSR (marked by the black box in panel (b)). The elevation profile comparison between ELunarDTMNet and LOLA DTMs is shown, with the profile location indicated by the yellow line in panel (c).

## 4. Discussion

### 4.1. New opportunities for advancing lunar science through high-resolution DTMs

Topographic models provide a critical foundation for lunar exploration and scientific studies; however, many of the available terrain data, such as SLDEM (Barker et al., 2016), LOLA DTM (Smith et al., 2017), and GLD100 (Scholten et al., 2012), are of relatively low resolution. This limitation motivates the development of high-resolution

DTMs, which can reduce uncertainties in quantitative studies of both lunar surface processes and the Moon's internal structure, while also providing more reliable topographic information for exploration site assessments. The extensive high-resolution imaging of the lunar surface by instruments such as LRO NAC and ShadowCam underscores the importance of efficient topographic modeling approaches for generating high-fidelity DTMs. As demonstrated in this study and by Chen et al. (2024), ELunarDTMNet was trained using SPG-derived DTMs, which provide geometric priors based on parallax between multiple viewpoints rather than brightness gradients, allowing the network to learn constraints that are largely independent of illuminations. Furthermore, ELunarDTMNet was designed not only to learn image brightness, but also to capture local and global context to better model the relationship between image data and topography. Therefore, ELunarDTMNet demonstrates robust performance under various solar azimuth and incidence angles, which allows a larger number of NAC images to be used for high-quality topographic reconstruction. In addition, while the SPG-derived DTMs are spatially sparse across the lunar surface, they capture a wide range of representative lunar surface features; as a result, ELunarDTMNet exhibits reliable generalization across terrain features of different scales, morphologies, and degradation states. These methodological advances open new opportunities for advancing lunar science by enabling high-resolution, large-scale topographic analyses that were previously limited by data availability and modeling efficiency.

**4.2. Unique advantages of high-quality mapping for the south polar region**

The persistently low solar elevation angles near the poles result in extensive shadowed areas, posing challenges for single-view topographic modeling using LRO NAC images, which rely on sufficient radiometric and textural information to infer surface topography (Chen et al., 2022). Following the launch of KPLO, high-resolution ShadowCam imagery directly addresses this limitation by providing usable imaging information within shadowed regions, thereby enabling detailed DTM reconstruction and facilitating comprehensive polar topographic analysis (Robinson et al., 2023). In addition, the polar regions exhibit several inherent characteristics that are favorable for high-quality topographic modeling with DL techniques. Firstly, the consistently low solar elevation angles in the polar regions result in imagery acquired under relatively stable illumination conditions, in contrast to low-latitude areas where large variations in solar incidence and viewing geometry need to be learned and accommodated by DL networks. Under such low and stable illumination conditions, as illustrated in Sect. 3.1, low solar elevation strengthens the coupling between surface relief and shading while reducing illumination-driven radiometric variability unrelated to topography, thereby preserving surface cues that are critical for topographic inference. Moreover, the polar orbit of LRO results in denser LOLA ground tracks at high latitudes, yielding LOLA DTMs with improved spatial coverage and sampling geometry (e.g., ~20 m resolution poleward of 80°S), in contrast to the widely used ~60 m SLDEM, which provides high-quality topography primarily at low latitudes. This provides a robust elevation reference for DL-based reconstruction, with two key roles: one is to supply low-frequency topographic constraints that guide high-resolution terrain recovery from imagery; the other is to serve as an absolute reference for converting predicted DTMs from normalized values to physical elevations. Overall, these characteristics establish a favorable framework for high-quality topographic modeling of the lunar south polar region. As demonstrated in Table 1 and Figs. 2 and 6, the RMSE of polar DTMs with respect to reference models is generally lower than that of DTMs derived for low-latitude regions.

**4.3. Application to topographic modeling of other large planetary bodies**

Beyond the Moon, DL-based single-view topographic modeling appears promising for efficient high-resolution 3D reconstruction on other planetary bodies (Chen et al., 2021; La Grassa et al., 2022). Among planetary bodies in the Solar System, the Moon offers a particularly favorable setting for DL-based topographic modeling, owing to its near-global coverage of meter-scale optical imagery and relatively dense laser altimetry measurements. Mars represents the next most data-rich target. Optical imagery from the High Resolution Imaging Science Experiment (HiRISE) enables submeter- to meter-scale surface observation, while the Mars Orbiter Laser Altimeter (MOLA) provides a globally consistent but relatively coarse elevation reference (Smith et al., 2001; McEwen et al., 2007). For example, the standard MOLA gridded DTMs typically have horizontal sampling on the order of ~460-1850 m, an order of magnitude coarser than the ~60 m resolution SLDEM and the 20 m resolution polar LOLA DTM for the Moon (Lin et al., 2010). Other low-resolution DTMs derived from the High Resolution Stereo Camera (HRSC) and the Context Camera (CTX) stereo models, with typical horizontal resolution of ~50-150 m and ~18-24 m, respectively, can serve as elevation constraints for DL-based topographic modeling, analogous to the role of LOLA on the Moon (Gwinner et al., 2010; Tao et al., 2018). While these datasets do not provide global coverage like MOLA, they offer broader spatial context than individual HiRISE observations and can help guide the reconstruction of high-resolution topography by providing reliable low-frequency elevation information across regional scales. Moreover, DL-based topographic modeling on Mars faces additional challenges compared to the Moon due to the presence of a thin but dynamic atmosphere, dust storms, seasonal changes, and more pronounced surface albedo variations (Wellington & Bell III, 2020; He et al., 2024 López-Cayuela et al., 2024). These factors introduce radiometric and morphological variability in optical imagery, necessitating larger and more diverse datasets for effective training and robust generalization, whereas the Moon's airless and relatively static surface allows for more straightforward topographic reconstruction. Mercury shares similarities with the Moon, such as an airless and relatively stable surface, which in principle favors DL-based topographic modeling.

However, the available datasets for Mercury have historically been limited: previous topographic models relied primarily on MESSENGER stereo imagery and sparse laser altimetry, resulting in DTMs with restricted spatial resolution and coverage (Solomon et al., 2007). The ongoing BepiColombo mission is expected to improve this situation by providing systematic laser altimetry from the BepiColombo Laser Altimeter (BELA) and new optical imaging from the Spectrometers and Imagers for MPO BepiColombo Integrated Observatory SYStem (SIMBIO-SYS) instrument suite (Benkhoff et al., 2021). Despite these improvements, the availability of high-resolution optical imagery on Mercury remains limited compared to the Moon, which constrains the direct application of DL framework trained from high-resolution images. Consequently, DL methods need to be adapted to account for the spatial and resolution limitations of Mercury's data.

## 5. Conclusion

Despite the availability of meter-level lunar imaging data, such as LRO NAC observations, their full potential for high-resolution topographic modeling has remained underexplored. DL-based single-view DTM reconstruction can overcome several limitations of conventional methods, such as the strict requirement for stereo pairs in SPG, while also enabling rapid and flexible topographic modeling from single images. Nevertheless, the performance of such methods has not yet been sufficiently validated, leading to a limited understanding and adoption within the planetary science community compared with traditional approaches. In this study, we build on an improved ELunarDTMNet to evaluate its robustness under varying illumination conditions and its generalization capability across different terrain types, as both are fundamental to its reliability. The results demonstrate that ELunarDTMNet can generate more consistent and higher-quality topographic models than single-view SFS methods under varying solar azimuths and elevations. It is capable of handling a wide range of topographic modeling tasks, from large-area continuous and seamless reconstruction to the recovery of fine-scale subtle features, and can faithfully model landforms that have formed during different geological periods. Furthermore, it can reconstruct high-quality DTMs for the lunar south polar region under low-solar illumination environments, including PSRs based on the ShadowCam images. These results suggest that DL-based single-view DTM reconstruction provides an effective way to fully exploit high-resolution lunar imaging data, enabling detailed and high-quality topographic modeling. This capability has the potential to substantially reduce the current reliance on low-resolution DTMs in lunar exploration and geological interpretation.